\documentclass[preprint,prc,showpacs,showkeys]{revtex4}
\usepackage{amsfonts}
\usepackage{amsmath}
\usepackage{amssymb}
\usepackage{graphicx}
\usepackage{rotating}

\providecommand{\U}[1]{\protect\rule{.1in}{.1in}}
\providecommand{\U}[1]{\protect\rule{.1in}{.1in}}
\providecommand{\U}[1]{\protect\rule{.1in}{.1in}}
\providecommand{\U}[1]{\protect\rule{.1in}{.1in}}

\begin{document}

\preprint{}
\title{Nuclear processes in solids: basic 2nd-order processes\\
}
\author{P\'{e}ter K\'{a}lm\'{a}n\footnote{%
retired from Budapest University of Technology and Economics, Institute of
Physics, \newline
e-mail: kalmanpeter3@gmail.com}}
\author{Tam\'{a}s Keszthelyi}
\affiliation{Budapest University of Technology and Economics, Institute of Physics,
Budafoki \'{u}t 8. F., H-1521 Budapest, Hungary\ }

\begin{abstract}
Nuclear processes in solid environment are investigated. It is shown that if
a slow, quasi-free heavy particle of positive charge interacts with a "free"
electron of a metallic host, it can obtain such a great magnitude of
momentum in its intermediate state that the probability of its nuclear
reaction with an other positively charged, slow, heavy particle can
significantly increase. It is also shown that if a quasi-free heavy particle
of positive charge of intermediately low energy interacts with a heavy
particle of positive charge of the solid host, it can obtain much greater
momentum relative to the former case in the intermediate state and
consequently, the probability of a nuclear reaction with a positively
charged, heavy particle can even more increase. This mechanism opens the
door to a great variety of nuclear processes which up till know are thought
to have negligible rate at low energies. Low energy nuclear reactions
allowed by the Coulomb assistance of heavy charged particles is partly
overviewed. Nuclear $pd$ and $dd$ reactions are investigated numerically. It
was found that the leading channel in all the discussed charged particle
assisted $dd$ reactions is the electron assisted $d+d\rightarrow $ $^{4}He$
process.
\end{abstract}

\volumenumber{number}
\issuenumber{number}
\eid{identifier}
\date[Date text]{date}
\received[Received text]{date}
\revised[Revised text]{date}
\accepted[Accepted text]{date}
\published[Published text]{date}
\startpage{1}
\endpage{}
\maketitle

\section{Introduction}

It is a standard of nuclear physics that heavy, charged particles $j$ and $k$
of like positive charge of charge numbers $z_{j}$ and $z_{k}$ need
considerable amount of relative kinetic energy $E$ determined by the height
of the Coulomb barrier in order to let the probability of a nuclear
interaction have significant value. Mathematically it appears in the energy
dependence of the cross section $\left( \sigma \right) $ of the
charged-particle induced reactions as 
\begin{equation}
\sigma \left( E\right) =S\left( E\right) \exp \left[ -2\pi \eta _{jk}\left(
E\right) \right] /E,  \label{sigma}
\end{equation}%
where $S\left( E\right) $ is the astrophysical factor, which can be written
as $S(E)=S(0)+S_{1}E+S_{2}E^{2}$, \cite{Angulo}. The Sommerfeld parameter 
\begin{equation}
\eta _{jk}\left( k\right) =z_{j}z_{k}\alpha _{f}\frac{\mu _{jk}c}{\hbar k},
\label{eta23}
\end{equation}%
where $k=\left\vert \mathbf{k}_{j}-\mathbf{k}_{k}\right\vert $ is the
magnitude of the relative wave vector $\mathbf{k=k}_{j}-\mathbf{k}_{k}$ of
the interacting particles of wave vectors $\mathbf{k}_{j}$ and $\mathbf{k}%
_{k}$, ($k\sim \sqrt{E}$). The reduced mass of particles $j$ and $k$ of rest
masses $m_{j}$ and $m_{k}$ 
\begin{equation}
\mu _{jk}=m_{j}m_{k}/\left( m_{j}+m_{k}\right) ,  \label{redukalt}
\end{equation}%
$\hbar $ is the reduced Planck constant, $e$ is the elementary charge and $%
\alpha _{f}$ is the fine structure constant. (It can be shown that in the
case of slow relative motion the exponential function in $\left( \ref{sigma}%
\right) $ is the same as the Gamow factor \cite{Gamow}, that hinders nuclear
reactions between particles of like electric charge.) The energy dependence
of the cross section $\left( \ref{sigma}\right) $ can be\ derived applying
the Coulomb solution 
\begin{equation}
\varphi (\mathbf{r})=e^{i\mathbf{k}\cdot \mathbf{r}}f(\mathbf{k,r})/\sqrt{V},
\label{Cb1}
\end{equation}%
which is the wave function of a free particle of charge number $z_{j}$ in a
repulsive Coulomb field of charge number $z_{k}$, where $V$ denotes the
volume of normalization, and $\mathbf{r}$ is the relative coordinate of the
two particles. Here 
\begin{equation}
f(\mathbf{k},\mathbf{r})=e^{-\pi \eta _{jk}/2}\Gamma (1+i\eta
_{jk})_{1}F_{1}(-i\eta _{jk},1;i[kr-\mathbf{k}\cdot \mathbf{r}]),
\label{Hyperg}
\end{equation}%
where $_{1}F_{1}$ is the confluent hypergeometric function and $\Gamma $ is
the Gamma function \cite{Alder}.

It is the consequence of energy dependence $\left( \ref{sigma}\right) $ of
the cross section that to this day it is a commonplace that the rate of any
nuclear reaction between heavy, charged particles of positive charge is
unobservable at low energies. The aim of this paper is to show that in a
solid (particularly in a metal), contrary to the former assumption, there
are nuclear processes that can have observable rate at low energies.

\section{Preliminary considerations}

In the low-energy range ($kR\ll 1$, where $R$ is the radius of a nucleon)
and for $\left\vert \mathbf{r}\right\vert \leq R$ the long wavelength
approximation 
\begin{equation}
\left\vert \varphi (\mathbf{r})\right\vert =\left\vert \varphi (\mathbf{0}%
)\right\vert =f_{jk}(k)/\sqrt{V}  \label{Cb2}
\end{equation}%
is valid, where%
\begin{equation}
f_{jk}(k)=\left\vert f(\mathbf{k},\mathbf{0})\right\vert =\sqrt{\frac{2\pi
\eta _{jk}\left( k\right) }{\exp \left[ 2\pi \eta _{jk}\left( k\right) %
\right] -1}}  \label{fjk}
\end{equation}%
is the Coulomb factor. We introduce the notation 
\begin{equation}
F_{jk}(k)=f_{jk}^{2\text{ }}(k)  \label{Fjk}
\end{equation}%
with which the cross section (and the rate) of a first order process is
proportional. If $k$ is small then the long wavelength approximation
produces the form $\left( \ref{sigma}\right) $ of $\sigma $ well. Thus the
fact that the rate of any nuclear reaction between heavy, charged particles
of positive charge is unobservable at low energies is the consequence of $%
F_{jk}(k)$ being small.

In solids, however, where free electrons are present nuclei also Coulomb
interact with electrons. It means that the Hamiltonian governing the state
of the nuclei must also contain the interaction Hamiltonian with the
electrons. If one wants to describe the interaction (scattering) of a
nucleus with an other one that takes into consideration the interaction with
the electrons the lower order process appears in the perturbation
calculation can be seen in FIG. 1. The basic idea can be demonstrated with
the aid of FIG.1(a), in which a Coulomb scattering is followed by a capture
process governed by strong interaction. When calculating the transition
probability (and the rate) of such a second order process the following
statements are valid. Energy and momentum (wave number vector) are
conserved, i.e. $E_{i}=E_{f}$, $\mathbf{k}_{i}=\mathbf{k}_{f}$, where $E_{i}$
and $E_{f}$ are the total energies, and $\mathbf{k}_{i}$ and $\mathbf{k}_{f}$
are the total wave number vectors in the initial and final states,
respectively. However energy-wave number vector (momentum) conservation may
be violated in the "intermediate" state. In the cases investigated the
initial particles (particles $1$, $2$ and $3$) are slow and the sum of their
initial kinetic energies $E_{i}$ and the sum of their wave number vectors $%
\mathbf{k}_{i}$ can be neglected, i.e. $E_{i}=0$ and $\mathbf{k}_{i}=0$ can
be supposed.

It is thought that particle $1$ is an electron and particles $2$ and $3$ are
heavy, and of positive charge. The nuclear reaction $2+3$ $\rightarrow 4$
has reaction energy $\Delta $. This energy is shared between the outgoing
particles $1$ and $4$. Thus particle $1$ obtains energy and wave number
vector of nuclear order of magnitude. Since the Coulomb interaction in the
case of free particles conserves wave number vector (momentum), and since
the initial wave number vector of particles $1$ and $2$ can be neglected in
wave number conservation, particle $2^{\prime }$ gets a wave number vector $%
\mathbf{k}_{2^{\prime }}$ opposite to the final wave number vector $\mathbf{k%
}_{1^{\prime }}$ of particle $1^{\prime }$, i.e. $\mathbf{k}_{2^{\prime }}=-%
\mathbf{k}_{1^{\prime }}$. Moreover if one calculates the Coulomb matrix
element using plane waves for the free particles then the matrix element
must be corrected with the so called Sommerfeld factor \cite{Heitler} 
\begin{equation}
g_{S}=\frac{f_{12}(\left\vert \mathbf{k}_{2}-\mathbf{k}_{1}\right\vert )}{%
f_{12}(\left\vert \mathbf{k}_{2^{\prime }}-\mathbf{k}_{1^{\prime
}}\right\vert )},  \label{gs}
\end{equation}%
where $f_{12}$-s are Coulomb factors [see $\left( \ref{fjk}\right) $] for
particles $1$ and $2$ of electric charge numbers $z_{2}=1$ and $z_{1}=-1$
since particle 1 is an electron. If particle $2$ is heavy and slow, and
particle $1$ is an electron then $\left\vert \mathbf{k}_{2}-\mathbf{k}%
_{1}\right\vert =k_{1}$, that is the magnitude of the initial wave number
vector of particle $1$, furthermore $\left\vert \mathbf{k}_{2^{\prime }}-%
\mathbf{k}_{1^{\prime }}\right\vert =2k_{2^{\prime }}$ since $\mathbf{k}%
_{2^{\prime }}=-\mathbf{k}_{1^{\prime }}$. Thus the cross section and the
rate are proportional to 
\begin{equation}
G_{S}\left( k_{1},2k_{2^{\prime }}\right) =g_{S}^{2}=\frac{F_{12}(k_{1})}{%
F_{12}(2k_{2^{\prime }})}.  \label{Somfact}
\end{equation}%
It can be shown that $F_{12}(2k_{2^{\prime }})=1$ and therefore $%
G_{S}=F_{12}(k_{1})=23.18\times \left( E_{1}^{-1/2}\left( eV\right) \right) $%
, where $E_{1}$ is the energy of the initial free electron. When calculating
the matrix element of the strong interaction potential between particles $2$
and $3$ we use $\left( \ref{Cb2}\right) $. Consequently the second order
rate is proportional to $G_{S}F_{23}$ which is mainly determined by $%
F_{23}(k_{2^{\prime }})\simeq \exp \left[ -2\pi \eta _{23}\left(
k_{2^{\prime }}\right) \right] $. Now $k_{2^{\prime }}=\left\vert \mathbf{k}%
_{1^{\prime }}\right\vert $ has a nuclear order of magnitude.

The first order rate is proportional to $F_{23}(k)\simeq \exp \left[ -2\pi
\eta _{23}\left( k\right) \right] $, where $k$ is the wave number of the
slow, initial particle $2$ (particle $3$ is supposed to beat rest). Since $%
k\ll k_{2^{\prime }}$, and therefore $\eta _{23}\left( k\right) \gg \eta
_{23}\left( k_{2^{\prime }}\right) $, and consequently $\exp \left[ -2\pi
\eta _{23}\left( k\right) \right] \ll \exp \left[ -2\pi \eta _{23}\left(
k_{2^{\prime }}\right) \right] $, the rate of the second order process is
much higher than that of the first order process. As a result, although the
rate of a second order process is usually much less than the rate of a first
order process, in this case the exponential increment is so huge that it can
dwarf the rate of the first order process.

As a numerical example we consider the electron assisted $d+d\rightarrow $ $%
_{2}^{4}He$ process with slow deuterons. In this case, one of the slow
deuterons (as particle $2$) can enter into Coulomb interaction with a
quasi-free, slow electron of the solid before the nuclear reaction (see FIG.
1(a)). The states of the free deuteron and the free electron can be
described by plane waves, therefore the Coulomb interaction preserves the
wave number vector (momentum). If in the second order process the Coulomb
interaction is followed by strong interaction, which induces a nuclear
capture process, then the energy $\Delta $ of the nuclear reaction is
divided between the electron and the heavy nuclear product. Since the rest
mass $m_{N}$ of the nuclear product is much larger than the rest mass $m_{e}$
of the electron, the electron will take almost all the total nuclear
reaction energy $\Delta $ away and the magnitude of its wave number vector $%
k_{1^{\prime }}=\sqrt{\Delta ^{2}+2m_{e}c^{2}\Delta }/\left( \hbar c\right)
\simeq \Delta /\left( \hbar c\right) $ $\left( \text{if }\Delta \gg
m_{e}c^{2}\right) $. If initially (before the Coulomb interaction) the
electron and the deuteron move slowly and the magnitudes of their wave
number vectors are much smaller than $\Delta /\left( \hbar c\right) $, then
the initial wave number vectors can be neglected in the wave number vector
(momentum) conservation and consequently, in the intermediate state the
deuteron gets a wave number vector of magnitude $k_{1^{\prime }}$ and of
direction opposite to the wave number vector of the electron. Thus the
deuteron will have a (virtual) wave number vector in the intermediate state
that is large enough to make it able to overcome the Coulomb barrier by
tunneling and to take part in a nuclear process. If $\Delta =23.84$ $\left[
MeV\right] $, which is the reaction energy of the $d+d\rightarrow $ $%
_{2}^{4}He$ reaction, then the deuteron will have a (virtual) wave number $%
k_{2^{\prime }}=\Delta /\left( \hbar c\right) $ in the intermediate state
(in state 2'). The corresponding value $F_{23}=0.356$. It must be compared
e.g. to the extremely small value $F_{23}\left( 1\text{ }eV\right)
=1.1\times 10^{-427}$ that is characteristic of the first order process.

\section{Electron assisted nuclear processes}

The change of state of heavy charged particles induced by solid state
environment is modelled in the following way. Let us take two independent
systems $A$ and $B$, where $A$ is a solid and $B$ is an ensemble of free,
heavy charged particles (e.g. a free deuteron or proton gas) with the
corresponding Hamiltonians $H_{A}$ and $H_{B}$. It is supposed that their
eigenvalue problems are solved, and the complete set of the eigenvectors of
the two independent systems are known. Let us extend the state vectors of
systems $A$ and $B$ to those nuclear bound states, which are initially
empty, corresponding to the assumption that at the beginning the two systems
do not interact. The interaction between them to be switched on
adiabatically is described by the interaction Hamiltonian $%
V_{AB}=V^{Cb}\left( \mathbf{x}_{AB}\right) +V^{St}\left( \mathbf{x}%
_{AB}\right) $, where $V^{Cb}$ and $V^{St}$ stand for the Coulomb and the
strong interaction potentials, respectively, and the suffixes $A$ and $B$ in
their argument symbolize that one party of the interaction comes from system 
$A$ and the other from system $B$. (Similar model is used by \cite{Louisell}
introducing the reduced density operator.) In the process investigated,
first a heavy, charged particle of system $B$ takes part in a Coulomb
scattering with any charged particle of system $A$ and it is followed by a
strong interaction with some nucleus of system $A$ that leads to their final
bound states. The graphs of the process can be seen in FIG. 1. The process
of FIG. 1(a) is a nuclear capture process and the process of FIG. 1(b) is a
nuclear reaction. (The processes, where the nuclear interaction is followed
by the Coulomb interaction, can be neglected because of the situation
discussed in the introduction.) Particles 1 and 3, belonging to system $A$
are: electrons (e.g. free electrons in the case of a metal) and localized
heavy, charged particles (bound, localized $p$, $d$ and other nuclei) as
nuclear targets. Particle 2 belongs to system $B$, that is a charged, heavy
particle (e.g. proton $\left( p\right) $ or deuteron $\left( d\right) $),
and that is supposed to move freely in a solid (e.g. in a metal). Since the
aim of this paper is to show the fundamentals of the main effect, the
problem, that there may be identical, indistinguishable particles in systems 
$A$ and $B$ is not considered, the simplest description is chosen, and the
dynamic evolution of the number $N_{2}$ of particles 2 of system $B$ is not
investigated.

\begin{figure}[ptb]
\resizebox{6.0cm}{!}{\includegraphics*{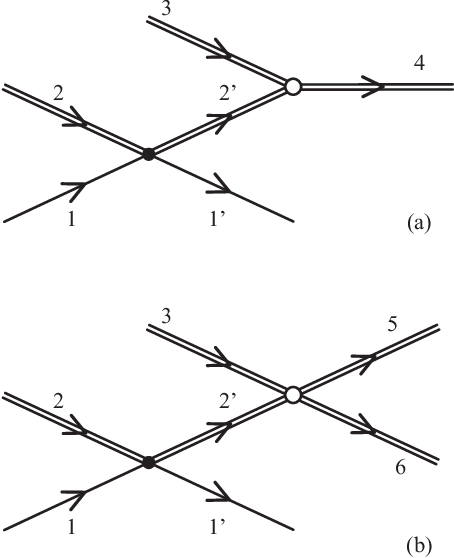}}
\caption{The graphs of electron assisted nuclear reactions. The simple lines
represent free (initial (1) and final (1')) electrons. The doubled lines
represent free, heavy, charged initial (2) particles (such as p, d), their
intermediate state (2'), target nuclei (3) and reaction products (4, 5, 6).
The filled dot denotes Coulomb-interaction and the open circle denotes
nuclear (strong) interaction. FIG. 1(a) is a capture process and FIG. 1(b)
is a reaction.}
\label{figure1}
\end{figure}

The transition probability per unit time $\left( W_{fi}^{(2)}\right) $ of
the process can be written as%
\begin{equation}
W_{fi}^{(2)}=\frac{2\pi }{\hbar }\sum_{f}\left\vert T_{if}^{(2)}\right\vert
^{2}\delta (E_{f}-\Delta )  \label{Wfie}
\end{equation}%
with 
\begin{equation}
T_{if}^{(2)}=\sum_{\mu }\frac{V_{f\mu }^{St}V_{\mu i}^{Cb}}{E_{\mu }-E_{i}}%
\frac{\left( 2\pi \right) ^{3}}{V}\delta \left( \mathbf{k}_{1f}+\mathbf{K}%
_{\left( type\right) }\right) ,  \label{Tif}
\end{equation}%
where $E_{i}$, $E_{\mu }$ and $E_{f}$ are the kinetic energies in the
initial, intermediate and final states, respectively, $\Delta $ is the
reaction energy, i.e. the difference between the rest energies of the
initial and final states,~$V$ is the volume of normalization. $V_{\mu
i}^{Cb} $ \ is the matrix element of the Coulomb potential between the
initial and intermediate states and $V_{f\mu }^{St}$ is the matrix element
of the potential of the strong interaction between intermediate and final
states. $type=\alpha ,+\beta ,n\beta $ correspond to the reaction of FIG.
1(a) $\left( \alpha \right) $ and to the reactions of\ FIG. 1(b) with both
particles charged $\left( +\beta \right) $ and with one of them neutral $%
\left( n\beta \right) $, respectively. 
\begin{equation}
E_{f}=E_{1f}+E_{\left( type\right) },  \label{Ef}
\end{equation}

\begin{equation}
E_{\mu }=E_{1f}\left( \mathbf{k}_{1f}\right) +E_{2}\left( \mathbf{k}_{2\mu
}\right) ,  \label{Em}
\end{equation}%
where $E_{1f}$ is the kinetic energy and $\mathbf{k}_{1f}$ is the wave
vector of particle 1 (electron) in the final state. $E_{\left( \alpha
\right) }=E_{4}\left( \mathbf{k}_{4}\right) $, $\mathbf{K}_{\left( \alpha
\right) }=\mathbf{k}_{4}$ in the case of process of FIG. 1(a) and $E_{\left(
+\beta \text{ }or\text{ }n\beta \right) }=E_{5}\left( \mathbf{k}_{5}\right)
+E_{6}\left( \mathbf{k}_{6}\right) $, $\mathbf{K}_{\left( +\beta \text{ }or%
\text{ }n\beta \right) }=\mathbf{k}_{5}+\mathbf{k}_{6}$ in the case of
process of FIG. 1(b) with $E_{j}\left( \mathbf{k}_{j}\right) $ the kinetic
energy of the $j-th$ particle in the intermediate (particle 2) and final
(particles 4 or 5, 6) states. Since particle 1 is an electron,%
\begin{equation}
E_{1f}=\sqrt{\left( \hbar c\right) ^{2}k_{1f}^{2}+m_{e}^{2}c^{4}}-m_{e}c^{2}
\label{E1f}
\end{equation}%
with $m_{e}c^{2}$ denoting the rest energy of the electron.

For the Coulomb potential we use its screened form%
\begin{equation}
V^{Cb}\left( \mathbf{x}\right) =\frac{e^{2}z_{1}z_{2}}{2\pi ^{2}}\int \frac{1%
}{q^{2}+\lambda ^{2}}\exp \left( i\mathbf{q}\cdot \mathbf{x}\right) d\mathbf{%
q}  \label{Vcb1}
\end{equation}%
with screening parameter $\lambda $ and coupling strength $e^{2}=\alpha
_{f}\hbar c$. For the strong interaction the interaction potential 
\begin{equation}
V^{St}\left( \mathbf{x}\right) =-2f^{2}\frac{\exp \left( -s\left\vert 
\mathbf{x}\right\vert \right) }{\left\vert \mathbf{x}\right\vert }
\label{VSt1}
\end{equation}%
is applied, where the strong coupling strength $f^{2}=0.08\hbar c$ \cite%
{Bjorken} and $1/s$ is the range of the strong interaction. The calculation
of the total rates of the electron assisted nuclear processes can be found
in Appendix I.

The result of the total rate of the leading, electron assisted ($\mathit{p}$
or $d$) capture process 
\begin{equation}
W_{tot}^{(2)}\left( \alpha \right) =K_{tot}\left( \alpha \right)
\left\langle G_{S}(k_{1i},2k_{1f})\right\rangle _{av}\frac{F_{23}(k_{1f})}{%
\Delta ^{4}}h_{corr,3}^{2}uN_{2}  \label{Wtot2alfa}
\end{equation}%
with $k_{1f}=\frac{\Delta }{\hbar c}$. Here $u$ denotes the deuteron (or
proton) over metal number densities, $N_{2}$ is the number of initial
particles 2 and for $G_{S}$ see $\left( \ref{Somfact}\right) $. Furthermore,
we introduced the following notation%
\begin{equation}
K_{tot}\left( \alpha \right) =\frac{32}{d^{6}}g_{e}K_{0}\left( \alpha
\right) ,  \label{Ktotalfa0}
\end{equation}%
where $g_{e}$ is the number of the valence electron states corresponding to
one unit cell and%
\begin{equation}
K_{0}\left( \alpha \right) =192\left( 2\pi \right)
^{2}z_{1}^{2}z_{2}^{2}\left( 1-\frac{2}{e}\right) ^{2}\alpha _{f}^{2}\left( 
\frac{f^{2}}{\hbar c}\right) ^{2}\left( \hbar c\right) ^{4}Rc.
\label{K0alfa}
\end{equation}%
Here $R=1.2\times 10^{-13}$ $\left[ cm\right] $ is the radius of a nucleon
(corresponding to the single nucleon approach applied). Finally, $%
h_{corr,3}=A_{3}-z_{3}$ in the case of proton capture process and $%
h_{corr,3}=A_{3}$ in the case of deuteron capture reactions (both are taken
in the Weisskopf approximation), where $A_{3}$ and $z_{3}$ are the mass and
charge numbers of particle 3.

Taking $z_{2}=1$, $\alpha _{f}=1/137$, $f^{2}/\left( \hbar c\right) =0.08$, $%
c=3\times 10^{10}$ $\left[ cm/s\right] $, $v_{c}=d^{3}/4$ (the volume of
unit cell for $Ni$ and $Pd$) with $d=3.89\times 10^{-8}$ $\left[ cm\right] $
($Pd$ lattice) one gets $K_{tot}\left( \alpha \right) =0.0091$ $\left[
MeV^{4}s^{-1}\right] $ in the case of $Pd$. Furthermore $\left\langle
G_{S}\right\rangle _{av}=\left\langle F_{12}(k_{1i})\right\rangle
_{av}=23.18\times \left\langle E_{1}^{-1/2}\left( eV\right) \right\rangle
_{av}$, where $E_{1}$ is the energy of the initial free electron and the
average is made by means of Fermi-Dirac distribution. ($N_{2}$ and $%
F_{23}(k_{1f}=\frac{\Delta }{\hbar c})$ are dimensionless and $\Delta $ has
to be substituted in $\left[ MeV\right] $ units in $\left( \ref{Wtot2alfa}%
\right) $.)

The quantity $F_{23}(k_{1f}=\frac{\Delta }{\hbar c})$ given by $\left( \ref%
{Fjk}\right) $ is the most important factor in $W_{tot}^{(2)}\left( \alpha
\right) $. The electron (particle 1) transfers large momentum (energy) to
particle 2' through Coulomb scattering. Consequently, particle 2' appears in
the nuclear process with the corresponding large wave number and the
probability of the nuclear process drastically increases. $F_{23}(\frac{%
\Delta }{\hbar c})$ has to be compared with $F_{23}(k_{i}(E_{i}))$, which is
the square of the Coulomb factor of the usual first order process, where $%
E_{i}$ is the kinetic energy in the relative motion of particles 2 and 3 in
the initial state of the usual, first order process. If $E_{i}$ has $eV$
order of magnitude then 
\begin{equation}
F_{23}(\frac{\Delta }{\hbar c})\gg F_{23}(k_{i}(E_{i})).
\label{FCbhrelation}
\end{equation}%
From this relationship it follows that although the usual first order $%
2+3\rightarrow 4$ nuclear process is very strongly hindered by the repulsive
Coulomb interaction due to the extremely small value of $%
F_{23}(k_{i}(E_{i})) $, in consequence of the appearance of the much larger
quantity $F_{23}(\frac{\Delta }{\hbar c})$ in the rate of the second order
electron assisted process the hindering effect practically disappears.

\section{Heavy particle assisted nuclear processes}

In electron assisted nuclear reactions heavy, charged particles are created
in the decelerating process of reaction products of the electron assisted
processes (e.g. $_{2}^{4}He$ of energy of $0.0758$ $MeV$ is created in the
electron assisted $d+d\rightarrow $ $_{2}^{4}He$ reaction). The energy of
these heavy particles may be intermediately low (of about $0.01$ $\left[ MeV%
\right] $) so their nuclear processes have to be also considered among the
accountable nuclear processes. The corresponding graphs can be seen in FIG.
2. The graph of FIG. 2(a) depicts a nuclear capture process and FIG. 2(b)
shows a nuclear reaction. (The processes, where the nuclear interaction is
followed by the Coulomb interaction, can also be neglected.) Now particles 1
and 3 belong to system $A$ and particle 2 which is a free, heavy particle
created in an electron assisted nuclear process belongs to system $B$.
According to the applied notation, particles 2 and 3 take part in a nuclear
process and particle 1 only assists it. The different processes will be
distinguished by the type of the assisting particle and also by the type of
the nuclear process. In our model charged, heavy particles, such as protons $%
\left( p\right) $, deuterons $\left( d\right) $ can form system $B$, they
may be particle 2, which are supposed to move freely in a solid (e.g. in a
metal). The particles, that take part in the processes and belong to system $%
A$ are: localized heavy, charged particles (bound, localized $p$, $d$ and
other nuclei) as the participants of Coulomb scattering (particle 1) and
localized heavy, charged particles (bound, localized $p$, $d$ and other
nuclei) as nuclear targets (particle 3). The problem, that there may be
identical particles in systems $A$ and $B$ that are indistinguishable, is
also disregarded here.

\begin{figure}[ptb]
\resizebox{6.0cm}{!}{\includegraphics*{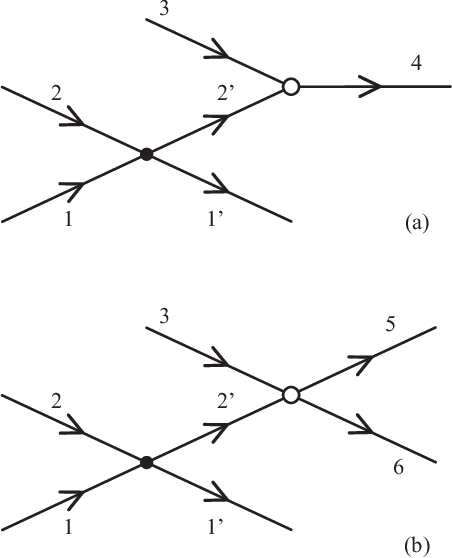}}
\caption{The graphs of heavy particle Coulomb-assisted nuclear processes.
The filled dot denotes Coulomb-interaction and the open circle denotes
nuclear (strong) interaction. The localized particles 1 and 3 belong to
system A. The free particle 2 (e.g. p or d) belongs to system B. All the
particles are heavy and positively charged. From the point of view of the
nuclear reaction particle 2' is ingoing, particle 3 is target and particles
4, 5, 6 are products of the process. FIG. 1(a) is a capture process and FIG.
1(b) is a reaction.}
\label{figure2}
\end{figure}

The calculation of the transition probability per unit time $\left(
W_{fi}^{(2),h}\right) $ of the process can be performed through similar
steps to those applied for the calculation of the rate of the electron
assisted process. The main difference is that particle 1 is heavy and
localized.

The matrix element of the screened Coulomb potential modifies as%
\begin{equation}
V_{\mu i}^{Cb}=\frac{4\pi e^{2}z_{1}z_{2}}{V^{3/2}}\frac{\widetilde{\psi }%
_{1i}\left( \mathbf{k}_{2i}-\mathbf{k}_{1f}-\mathbf{k}_{2\mu }\right) }{%
\left\vert \mathbf{k}_{2i}-\mathbf{k}_{2\mu }\right\vert ^{2}+\lambda ^{2}}%
g_{S}.  \label{VCbmi}
\end{equation}%
Here%
\begin{equation}
\widetilde{\psi }_{1i}\left( \mathbf{K}\right) =\int \psi _{1i}\left( 
\mathbf{x}\right) e^{-i\mathbf{K}\cdot \mathbf{x}}d\mathbf{x,}
\label{psziFour}
\end{equation}%
where $\psi _{1i}\left( \mathbf{x}\right) $ stands for the initial,
localized state of particle 1. It is supposed that

\begin{equation}
\psi _{1i}\left( \mathbf{x}\right) =\left( \frac{\beta _{1}^{2}}{\pi }%
\right) ^{3/4}e^{-\frac{\beta _{1}^{2}}{2}\mathbf{x}^{2}}  \label{dfi}
\end{equation}%
is the wave function of the ground state of a 3-dimensional harmonic
oscillator of angular frequency $\omega _{1}$ with $\beta _{1}=\sqrt{%
m_{1}\omega _{1}/\hbar }$ \cite{Cohen}. The calculation of the total rates
of the heavy particle assisted nuclear processes can be found in Appendix II.

The total rate of the leading, heavy particle assisted nuclear capture
processes valid in the case of $fcc$ metals 
\begin{align}
W_{tot}^{(2),h}\left( \alpha \right) & =\frac{8K_{0}^{h}\left( \alpha
\right) \beta _{1}^{3}}{d^{3}\sqrt{m_{0}c^{2}}}G_{S}\left(
k_{2i},2k_{1f}\right) h_{corr,3}^{2}\times \\
& \times \left( \frac{a_{12}}{a_{14}}\right) ^{2}\frac{\left(
z_{1}z_{2}\right) ^{2}}{\sqrt{a_{14}}}\frac{F_{23}^{h}(\Delta )}{\Delta
^{7/2}}uN_{2}.  \notag
\end{align}%
Here the parameter $\beta _{1}$, e.g. in the case of a localized deuteron,
is $\beta _{1}^{d}=\sqrt{m_{d}\omega /\hbar }$, where $m_{d}$ is the
deuteron rest mass, $\hbar \omega $ corresponds to the energy of an optical
phonon in the deuterized metal, and $d$ is the lattice parameter, 
\begin{equation}
K_{0}^{h}\left( \alpha \right) =K_{0}\left( \alpha \right) /\left( 2\pi
\right) ^{3/2}.  \label{K0halfa}
\end{equation}%
For $K_{0}\left( \alpha \right) $ see $\left( \ref{K0alfa}\right) $. The
quantity%
\begin{equation}
F_{23}^{h}(\Delta )=\frac{2\pi \eta _{23}^{h}\left( \Delta \right) }{\exp %
\left[ 2\pi \eta _{23}^{h}\left( \Delta \right) \right] -1}  \label{Fh230}
\end{equation}%
with the parameter 
\begin{equation}
\eta _{23}^{h}\left( \Delta \right) =z_{2}z_{3}\alpha _{f}a_{23}\sqrt{\frac{%
m_{0}c^{2}}{2a_{14}\Delta }},  \label{eta23h0}
\end{equation}%
where $m_{0}c^{2}=931.494$ $\left[ MeV\right] $ is the atomic mass unit, and 
\begin{equation}
a_{jk}=\frac{A_{j}A_{k}}{A_{j}+A_{k}}  \label{ajk}
\end{equation}%
is the reduced mass number of particles $j$ and $k$ of mass numbers $A_{j}$
and $A_{k}$.

It can be seen from $\left( \ref{Fh230}\right) $ and $\left( \ref{eta23h0}%
\right) $ that this process opens the door for a great variety of nuclear
processes. To get an order of magnitude estimation of the effect we take $%
A_{1}\simeq A_{3}\gg A_{2}$, thus $a_{23}=A_{2}$ and $a_{14}=A_{4}/2\simeq
A_{3}/2=z_{3}$. With these approximations 
\begin{equation}
2\pi \eta _{23,app}^{h}=2\pi z_{2}\alpha _{f}A_{2}\sqrt{\frac{z_{3}m_{0}c^{2}%
}{2\Delta _{app}}}.  \label{eta23happrox0}
\end{equation}%
We take as a typical value $\Delta _{app}=4$ $\left[ MeV\right] $ in $\left( %
\ref{eta23happrox0}\right) $ that yields $2\pi \eta
_{23,app}^{h}=0.497\times z_{2}A_{2}z_{3}^{1/2}$. If $2\pi \eta
_{23,app}^{h}\leq 15$, one can obtain $F_{23}^{h}(\Delta )\geq 4.59\times
10^{-5}$, that can be produced with any pair of heavy particles 2 and 3 of $%
z_{2}A_{2}z_{3}^{1/2}\leq 30$. From this condition one can draw an important
and surprising conclusion: it allows a great variety of nuclear processes,
that were thought improbable up till now. Moreover, in many cases $2\pi \eta
_{23}^{h}\ll 1$, consequently $F_{23}^{h}(\Delta )=1$ (see $\left( \ref%
{Fh230}\right) $) and the hindering role of $F_{23}^{h}(\Delta )$
disappears. In order to show the capability of the heavy particle assisted
nuclear processes, some cases of the proton assisted proton captures 
\begin{equation}
\text{ }_{Z}^{A}X+p\rightarrow \text{ }_{Z+1}^{A+1}Y+\Delta
\end{equation}%
are investigated in Appendix III.

\section{Charged particle assisted $dd$ and $pd$ fusion processes in solid
metals}

As an example, let us take the simplest charged particle processes, the
usual nuclear fusion processes:

\begin{equation}
d+d\rightarrow\text{ }^{4}He+\gamma+23.847\text{ }MeV,  \label{R1}
\end{equation}

\begin{equation}
d+d\rightarrow\text{ }p+\text{ }^{3}H+4.033\text{ }MeV,  \label{R2}
\end{equation}

\begin{equation}
d+d\rightarrow n+\text{ }^{3}He+3.269\text{ }MeV,  \label{R3}
\end{equation}

\begin{equation}
p+d\rightarrow \text{ }^{3}He+\gamma +5.493\text{ }MeV.  \label{R4}
\end{equation}%
The coefficients of the astrophysical factor (see Introduction) of these
nuclear fusion processes are available \cite{Angulo}. Owing to their
astrophysical importance the low energy range of the above fusion processes
has been extensively investigated. In the extremely low energy range, i.e.
at near room temperature, the $S(E)=S(0)$ approximation is valid and the
ratio of the rates of the processes $\left( \ref{R1}\right) $, $\left( \ref%
{R2}\right) $ and $\left( \ref{R3}\right) $ is determined by their $S(0)$
values and therefore it can be considered energy independent. The
corresponding $S(0)$ values of processes $\left( \ref{R1}\right) $, $\left( %
\ref{R2}\right) $, $\left( \ref{R3}\right) $ and $\left( \ref{R4}\right) $
are: $5.0\times 10^{-9}$, $5.6\times 10^{-2}$, $5.5\times 10^{-2}$ and $%
2.0\times 10^{-7}$ in $\left[ MeVb\right] $ units \cite{Angulo}.
Consequently, in a deuteron plasma or in a proton-deuteron mixed plasma the
leading processes are the processes $\left( \ref{R2}\right) $ and $\left( %
\ref{R3}\right) $ with approximately the same rates, and the rates of
processes $\left( \ref{R1}\right) $ and $\left( \ref{R4}\right) $ are many
orders of magnitude smaller.

For comparison we have calculated the $S(0)$ values of the above processes
in the simple nuclear model used in this article and supposing magnetic type
transition in reactions $\left( \ref{R1}\right) $ and $\left( \ref{R4}%
\right) $. The results are: $5.2\times 10^{-8}$, $5.0\times 10^{-3}$, $%
4.4\times 10^{-3}$ and $4.5\times 10^{-9}$ in $\left[ MeVb\right] $ units
corresponding to reactions $\left( \ref{R1}\right) $, $\left( \ref{R2}%
\right) $, $\left( \ref{R3}\right) $ and $\left( \ref{R4}\right) $,
respectively, indicating that our simple model is able to give qualitative
or semi quantitative conclusions.

Now we review the ratios of the rates of the processes obtained in solid
metals. In the case of the second order, simple electron assisted nuclear
processes we have obtained for the relevant quantities (of the corresponding
rates, see $\left( \ref{Wtot0}\right) $, $\left( \ref{Wtot+}\right) $ and $%
\left( \ref{Wtotn}\right) $) $F_{23}(\frac{\Delta }{\hbar c})=0.356$, $%
K_{2}K_{+}J_{22}=1.38\times 10^{-4}$ and $K_{2}J_{32}=5.79\times 10^{-4}$ \
for the electron assisted versions of processes $\left( \ref{R1}\right) $, $%
\left( \ref{R2}\right) $, $\left( \ref{R3}\right) $, respectively, and $%
F_{23}(\frac{\Delta }{\hbar c})=0.0282$\ for the electron assisted version
of process $\left( \ref{R4}\right) $ (see Appendix I.). From these numerical
values one can conclude that, contrary to the above, in the family of second
order electron assisted $dd$ processes the electron assisted version of
process $\left( \ref{R1}\right) $ is leading, and the electron assisted
versions of processes $\left( \ref{R2}\right) $ and $\left( \ref{R3}\right) $
have much lower rate. The total rate of the electron assisted version of $%
\left( \ref{R1}\right) $ is 
\begin{equation}
W_{tot}^{(2)}\left( \alpha \right) =2.0\times 10^{-7}\times \left\langle
E_{1}^{-1/2}\left[ eV\right] \right\rangle _{av}uN_{2}\left[ s^{-1}\right] ,
\label{WelassR1}
\end{equation}%
where $E_{1}$ is the energy of the initial free electron in the conduction
band. Averaging $E_{1}^{-1/2}$ by means of the Fermi-Dirac distribution in
the Sommerfeld free electron model at $T=0$ yields $\left\langle E_{1}^{-1/2}%
\left[ eV\right] \right\rangle _{av}=2\left( E_{F}^{-1/2}\left[ eV\right]
\right) $, where $E_{F}$ denotes the Fermi energy \cite{Solyom}.

In the case of heavy particle assisted processes we have obtained $%
F_{23}^{h}(\Delta )=0.902$, and $W_{tot}^{(2),h}\left( \alpha \right)
=4.2\times 10^{-8}\times G_{S}N_{2}u$ $\left[ s^{-1}\right] $\ for process $%
d+d\rightarrow $ $^{4}He$ in the case of a deuterized $Pd$ target, i.e. the
particles 1, 2 and 3 are all deuterons. In the case of process $%
p+d\rightarrow $ $^{3}He$ and in the same target material, i.e. particles 1
and 3 are deuterons, and particle 2 is a proton $F_{23}^{h}(\Delta )=0.846$
and $W_{tot}^{(2),h}\left( \alpha \right) =6.1\times 10^{-6}\times
G_{S}N_{2}u$ $\left[ s^{-1}\right] $.

From the above rates one can conclude that if energetic, heavy charged
particles are present in the sample the heavy particle assisted processes
are not negligible and among all the charged particle assisted processes
discussed here the electron assisted version of the $\left( \ref{R1}\right) $
process is the leading one.

\section{Summary}

It is found that, contrary to the commonly accepted opinion, in a solid
metal surrounding nuclear reactions can happen between heavy, charged
particles of like (positive) charge of low initial energy. It is recognized,
that one of the participant particles of a nuclear reaction of low initial
energy may pick up great momentum in a Coulomb scattering process on a free,
third particle of the surroundings. The virtually acquired great momentum,
that is determined by the energy of the reaction, can help to overcome the
hindering Coulomb barrier and can highly increase the rate of the nuclear
reaction even in cases when the rate would be otherwise negligible. It is
found that the electron assisted $d+d\rightarrow $ $^{4}He$ process has the
leading rate. In the reactions discussed energetic charged particles are
created, that can become (directly or after Coulomb collisions) the source
of heavy charged particles of intermediately low (of about a few $keV$)
energy. These heavy particles can assist nuclear reactions too. It is worth
mentioning that the shielding of the Coulomb potential has no effect on the
mechanisms discussed.

Our thoughts were motivated by our former theoretical findings \cite{kk2}
according to which the leading channel of the $p+d\rightarrow $ $^{3}He$
reaction in solid environment is the so called solid state internal
conversion process, an adapted version of ordinary internal conversion
process \cite{Hamilton}. In the process formerly discussed \cite{kk2} if the
reaction takes place in solid material, in which instead of the emission of
a $\gamma $ photon, the nuclear energy is taken away by an electron of the
environment (the metal), the Coulomb interaction induces a $p+d\rightarrow $ 
$^{3}He$ nuclear transition. The processes discussed here can be considered
as an alternative version of the solid state internal conversion process
since it is thought that one party of the initial particles of the nuclear
process takes part in Coulomb interaction with a charged particle of the
solid material (e.g. of a metal).

There may be many fields of physics where the traces of the proposed
mechanism may have been previously appeared. It is not the aim of this work
to give a systematic overview these fields. We only mention here two of them
that are thought to be partly related or explained by the processes
proposed. The first is the so called anomalous screening effect observed in
low energy accelerator physics investigating astrophysical factors of
nuclear reactions of low atomic numbers \cite{Raiola1}. The other one is the
family of low energy nuclear fusion processes. The physical background,
discussed in the Introduction and in the first part of Section V., was
questioned by the two decade old announcement \cite{FP1} on excess heat
generation due to nuclear fusion reaction of deuterons at deuterized Pd
cathodes during electrolysis at near room temperature. The paper \cite{FP1}
initiated continuous experimental work whose results were summarized
recently \cite{Storms2}. The mechanisms discussed here can explain some of
the main problems raised in \cite{Storms2}. (a) The mechanisms proposed here
make low energy fusion reactions and nuclear transmutations possible. (b)
The processes discussed explain the lack of the normally expected reaction
products.

The authors are indebted to K. H\"{a}rtlein for his technical assistance.

\section{Appendix I. - Rate calculation of electron assisted nuclear
processes}

For particles 1 and 2 (electron and ingoing heavy particle) taking part in
Coulomb interaction we use plane waves. Thus the Coulomb matrix element is
calculated in the Born approximation 
\begin{equation}
V_{\mu i}^{Cb}=\frac{2\left( 2\pi \right) ^{4}e^{2}z_{1}z_{2}}{V^{2}}\frac{%
\delta \left( \mathbf{k}_{2i}+\mathbf{k}_{1i}-\mathbf{k}_{1f}-\mathbf{k}%
_{2\mu }\right) }{\left\vert \mathbf{k}_{1f}-\mathbf{k}_{1i}\right\vert
^{2}+\lambda ^{2}}g_{S},  \label{VCbmuifree}
\end{equation}%
which is corrected with the so called Sommerfeld factor $g_{S}$ (see $\left( %
\ref{gs}\right) $). In the intermediate state we use plane waves of wave
vector $\mathbf{k}_{2\mu }$ for particle 2. The final state of the electron
is also a plane wave. The Dirac delta $\delta \left( \mathbf{k}_{2i}+\mathbf{%
k}_{1i}-\mathbf{k}_{1f}-\mathbf{k}_{2\mu }\right) $ will result $\mathbf{k}%
_{2\mu }=\mathbf{k}_{2i}+\mathbf{k}_{1i}-\mathbf{k}_{1f}$ in the
intermediate state (after integration over $\mathbf{k}_{2\mu }$, see later),
i.e. in the intermediate state particle 2' may have large momentum, which is
determined by the reaction energy due to the final wave vector of the
electron.

When calculating the matrix element of the strong interaction potential
between particles 2 and 3 we use the approximate form of $\varphi (\mathbf{r}%
)$ given in $\left( \ref{Cb2}\right) $.

For the process 1(a) the Weisskopf approximation is used, i.e. for the final
nuclear state of one nucleon (of particle 4) \ we take 
\begin{equation}
\Phi _{fW}\left( \mathbf{r}\right) =\sqrt{\frac{3}{4\pi R^{3}}}
\label{finagy}
\end{equation}%
if $r\leq R$, where $R$ is the nucleon radius, and $\Phi _{fW}\left( \mathbf{%
r}\right) =0$ for $r>R$. In evaluating $V_{f\mu }^{St}$ the long wavelength
approximation ($\exp \left( i\mathbf{k}_{2\mu }\cdot \mathbf{x}\right) =1$)
is used with $sR=1$ that results%
\begin{equation}
V_{f\mu }^{St,W}\left( \alpha \right) =-2f^{2}f_{23}(k_{2\mu })\sqrt{\frac{%
12\pi R}{V}}\left( 1-\frac{2}{e}\right)  \label{VStfm}
\end{equation}%
in single nucleon approach. In the case of process 1(b), for the final
states plane waves are assumed producing 
\begin{equation}
V_{f\mu }^{St,W}\left( \beta \right) =-2f^{2}f_{23}(k_{2\mu })\frac{4\pi
R^{2}}{V}  \label{VStfmb}
\end{equation}%
with a neutron as one of the particles 5 and 6. If both particles\ 5 and 6
have positive charge then the expression of $V_{f\mu }^{St,W}\left( \beta
\right) $ must be multiplied by $f_{56}$. The wave vectors $\mathbf{k}_{1i}$%
, $\mathbf{k}_{2i}$ of the particles 1 and 2 in their initial state can be
neglected in the calculation since $\left\vert \mathbf{k}_{1i}\right\vert
\ll \left\vert \mathbf{k}_{1f}\right\vert $, $\left\vert \mathbf{k}%
_{2i}\right\vert \ll \left\vert \mathbf{k}_{1f}\right\vert $, furthermore $%
\lambda \ll \left\vert \mathbf{k}_{1f}\right\vert $. Collecting everything
obtained above and applying the $\sum_{\mu }\rightarrow \int \left[ V/\left(
2\pi \right) ^{3}\right] d\mathbf{k}_{2\mu }$ correspondence%
\begin{align}
T_{if}^{(2)}(type)& =B(type)z_{1}z_{2}\delta \left( \mathbf{k}_{1f}+\mathbf{K%
}_{\left( type\right) }\right) \times  \label{TFi2} \\
& \int \frac{g_{S}f_{23}(k_{2\mu })}{\left( E_{\mu }-E_{i}\right) }\frac{%
\delta \left( \mathbf{k}_{1f}+\mathbf{k}_{2\mu }\right) }{\left\vert \mathbf{%
k}_{1f}\right\vert ^{2}}d\mathbf{k}_{2\mu }  \notag
\end{align}%
where

\begin{equation}
B(\alpha )=B_{0}\frac{\sqrt{3\pi R}}{V^{5/2}}\left( 1-\frac{2}{e}\right) ,
\label{Ba}
\end{equation}%
with 
\begin{equation*}
B_{0}=-8\left( 2\pi \right) ^{4}\alpha _{f}\left( \frac{f^{2}}{\hbar c}%
\right) \left( \hbar c\right) ^{2},
\end{equation*}%
\begin{equation}
B(n\beta )=B_{0}\frac{2\pi R^{2}}{V^{3}},  \label{Bnb}
\end{equation}%
and 
\begin{equation}
B(+\beta )=B(n\beta )f_{56}.  \label{B+b}
\end{equation}%
For more precise result, beyond the Weisskopf and long wavelength
approximations, and beyond the single nucleon approach the integrand of $%
\left( \ref{TFi2}\right) $ must be multiplied by a model dependent
correction factor 
\begin{equation}
h_{corr,k}=\frac{V_{f\mu }^{St}\left( type,\mathbf{k}_{\mu }\right) }{%
V_{f\mu }^{St,W}\left( type\right) },  \label{hcorr}
\end{equation}%
of the $k-th$ target particle, where $V_{f\mu }^{St}\left( type,\mathbf{k}%
_{\mu }\right) $ is the nuclear matrix element calculated in an other model,
without the long wavelength approximation and beyond the single nucleon
approach.

We will mainly treat proton (and deuteron) capture processes in which the
interaction of proton (deuteron) takes place with more than one nucleon. In
the Weisskopf approximation the sum of the matrix elements of proton-proton
and neutron-neutron interactions can be neglected due to the presence of
exchange terms, so in the case of proton capture the matrix element of the
strong interaction must be multiplied by the number of interacting neutrons
in the nucleus. In our case it means the neutron number of particle 3 $%
\widehat{N}_{3}=A_{3}-z_{3}$ ($h_{corr,3}=\widehat{N}_{3}$). In the case of
deuteron capture reactions and also in the Weisskopf approximation the
matrix-element must be multiplied by $A_{3}$ ($h_{corr,3}=A_{3}$).

\subsection{Rates of electron assisted $p$ or $d$ capture processes}

First the case $type=\alpha $ is treated. Neglecting the initial kinetic
energies and $E_{2\mu }$ in the denominator of $\left( \ref{Tif}\right) $, $%
E_{\mu }-E_{i}=E_{1f}+E_{2\mu }\simeq E_{1f}$ that will have a value $%
E_{1f}\simeq \Delta $ because of the energy Dirac delta. Using the $%
\sum_{\mu }\rightarrow \int \left[ V/\left( 2\pi \right) ^{3}\right] d%
\mathbf{k}_{2\mu }$ and the $\sum_{f}\rightarrow \int \left[ V/\left( 2\pi
\right) ^{3}\right] d\mathbf{k}_{4}\times \int \left[ V/\left( 2\pi \right)
^{3}\right] d\mathbf{k}_{1f}$ substitutions, the identities $\left[ \left(
2\pi \right) ^{3}\delta \left( \mathbf{k}_{4}+\mathbf{k}_{1f}\right) \right]
^{2}=\left( 2\pi \right) ^{3}\delta \left( \mathbf{k}_{4}+\mathbf{k}%
_{1f}\right) \left( 2\pi \right) ^{3}\delta \left( \mathbf{0}\right) $ and $%
\left( 2\pi \right) ^{3}\delta \left( \mathbf{0}\right) =V$, and the $%
k_{1f}dk_{1f}=E_{1f}dE_{1f}/\left( \hbar c\right) ^{2}$ relations,
furthermore carrying out integrations one can obtain%
\begin{equation}
W_{fi}^{(2)}\left( \alpha \right) =K_{0}\left( \alpha \right) G_{S}\left(
k_{1i},2k_{1f}\right) \frac{F_{23}(k_{1f})}{V^{2}\Delta ^{4}}h_{corr,3}^{2}.
\label{Wfi}
\end{equation}%
Here $G_{S}\left( k_{1i},2k_{1f}\right) =g_{S}^{2}$ [see $\left( \ref%
{Somfact}\right) $], remember that $F_{jk}(k)=f_{jk}^{2\text{ }}(k)$ [see $%
\left( \ref{Fjk}\right) $] and $K_{0}\left( \alpha \right) $ is defined in $%
\left( \ref{K0alfa}\right) $. In the case $E_{1f}\gg m_{e}c^{2}$ 
\begin{equation}
k_{1f}=\Delta /\left( \hbar c\right) .  \label{k1f}
\end{equation}

If one is interested in the total rate $W_{tot}^{(2)}\left( \alpha \right) $
of a sample then $W_{fi}^{(2)}$ must be multiplied by the numbers of initial
particles, $N_{1}$, $N_{2}$ and $N_{3}$ corresponding to electrons, and
particles 2 and 3, respectively. The quantity $N_{1}/V=g_{e}/v_{c}$, where $%
v_{c}$ is the volume of the elementary cell of the solid metal and $g_{e}$
is the number of the valence electron states corresponding to one unit cell,
e.g. $g_{e}=10$ and $v_{c}=d^{3}/4$ in the cases of $Ni$ and $Pd$ will be
discussed later. We introduce $N_{3}/V=n_{3}$, the number density of
particles 3. It is reasonable to take for the number density $n_{3}$ of the
target $n_{3}=2u/v_{c}=u8/d^{3}$ in the case of $fcc$ metals (such as $Pd$
and $Ni$) with $u$ denoting the deuteron (or proton) over metal number
densities. The result is 
\begin{equation}
W_{tot}^{(2)}\left( \alpha \right) =K_{tot}\left( \alpha \right)
\left\langle G_{S}(k_{1i},2k_{1f})\right\rangle _{av}\frac{F_{23}(\frac{%
\Delta }{\hbar c})}{\Delta ^{4}}h_{corr,3}^{2}uN_{2}  \label{Wtot0}
\end{equation}%
where $K_{tot}\left( \alpha \right) $ is determined by $\left( \ref%
{Ktotalfa0}\right) $ and the average is made by means of Fermi-Dirac
distribution.

\subsection{Rates of electron assisted processes with two outgoing heavy
particles}

In the case of $type=+\beta $ or $n\beta $ (for the process $\left( b\right) 
$ of FIG.1) the calculation is modified as follows. Let the wave vectors of
particles 5 and 6 be $\mathbf{k}_{5}$ and $\mathbf{k}_{6}$. This time, the $%
\sum_{f}\rightarrow \int \left[ V/\left( 2\pi \right) ^{3}\right] d\mathbf{k}%
_{5}\times \int \left[ V/\left( 2\pi \right) ^{3}\right] d\mathbf{k}%
_{6}\times \int \left[ V/\left( 2\pi \right) ^{3}\right] d\mathbf{k}_{1f}$
correspondence and the identities $\left[ \delta \left( \mathbf{k}_{1f}+%
\mathbf{k}_{5}+\mathbf{k}_{6}\right) \right] ^{2}=\delta \left( \mathbf{k}%
_{1f}+\mathbf{k}_{5}+\mathbf{k}_{6}\right) \delta \left( \mathbf{0}\right) $
and $\left( 2\pi \right) ^{3}\delta \left( \mathbf{0}\right) =V$ are used.
The integration over $\mathbf{k}_{1f}$ results $\mathbf{k}_{1f}=-\mathbf{k}%
_{5}-\mathbf{k}_{6}$. Using again the $N_{3}/V=n_{3}=2u/v_{c}=u8/d^{3}$
relation, 
\begin{equation}
W_{tot}^{(2)}\left( \beta \right) =K_{tot}\left( \alpha \right) \frac{%
\left\langle G_{S}\left( k_{1i},2k_{1f}\right) \right\rangle _{av}}{6\pi
^{2}\left( 1-\frac{2}{e}\right) ^{2}}\left( \frac{R}{\hbar c}\right) ^{3}%
\frac{J}{\Delta }h_{corr,3}^{2}uN_{2}  \label{Wtot2}
\end{equation}%
with%
\begin{align}
J& =\int \int \frac{\delta \left( \varepsilon \right) }{\left\vert \mathbf{%
\kappa }_{5}+\mathbf{\kappa }_{6}\right\vert ^{4}\varepsilon _{1f}^{2}\left( 
\mathbf{\kappa }_{5}+\mathbf{\kappa }_{6}\right) }\times  \label{i} \\
& \times F_{23}(\frac{\Delta }{\hbar c}\left\vert \mathbf{\kappa }_{5}+%
\mathbf{\kappa }_{6}\right\vert )F_{56}(\frac{\Delta }{\hbar c}\left\vert 
\mathbf{\kappa }_{5}-\mathbf{\kappa }_{6}\right\vert )d\mathbf{\kappa }_{5}d%
\mathbf{\kappa }_{6},  \notag
\end{align}%
where 
\begin{equation}
\delta \left( \varepsilon \right) =\delta \left[ \varepsilon _{5}\left( 
\mathbf{\kappa }_{5}\right) +\varepsilon _{6}\left( \mathbf{\kappa }%
_{6}\right) +\varepsilon _{1f}\left( \mathbf{\kappa }_{5}+\mathbf{\kappa }%
_{6}\right) -\delta _{e}-1\right] .  \label{deps}
\end{equation}%
Here $\mathbf{\kappa }_{j}=\hbar c\mathbf{k}_{j}/\Delta $ and $\varepsilon
_{j}=E_{j}/\Delta =\left[ \mathbf{\kappa }_{j}^{2}/\left( 2m_{j}\right) %
\right] \Delta $ are dimensionless quantities with $E_{j}$ as the kinetic
energy of particle $j$\ ($j$ $=5$ or $6$), $\delta _{e}=m_{e}c^{2}/\Delta $, 
$\varepsilon _{1f}=$ $E_{1f}/\Delta =\sqrt{\left\vert \mathbf{\kappa }_{5}+%
\mathbf{\kappa }_{6}\right\vert ^{2}+\delta _{e}^{2}}-\delta _{e}$ and the
suffixes $23$ of $F_{23}$ and $56$ of $F_{56}$ refer to particles 2, 3 and
5, 6, respectively. If one of the particles 5 and 6 is a neutron then%
\begin{equation}
F_{56}=1.  \label{fCb56n}
\end{equation}%
It is useful to introduce the following new variables in $\left( \ref{i}%
\right) $: 
\begin{equation}
\mathbf{a}=\mathbf{\kappa }_{5}+\mathbf{\kappa }_{6}
\end{equation}%
and%
\begin{equation}
\mathbf{b}=\mathbf{\kappa }_{5}-\mathbf{\kappa }_{6}.
\end{equation}%
Thus, if both particles 5 and 6 are positively charged then $J=J_{+\beta }$, 
\begin{equation}
J_{+\beta }=A_{23}A_{56}J_{2}  \label{JJ2}
\end{equation}%
where%
\begin{equation}
A_{23}=2\pi z_{2}z_{3}\alpha _{f}\mu _{23}c^{2}/\Delta ,  \label{A23}
\end{equation}%
\begin{equation}
A_{56}=2\pi z_{5}z_{6}\alpha _{f}\mu _{56}c^{2}/\Delta  \label{A56}
\end{equation}%
and 
\begin{align}
J_{2}& =\int \int \frac{\delta \left[ \varepsilon \left( \mathbf{a},\mathbf{b%
}\right) \right] }{\left\vert \mathbf{a}\right\vert ^{5}\left\vert \mathbf{b}%
\right\vert \varepsilon _{1f}^{2}\left( \mathbf{a}\right) }\times  \label{J2}
\\
& \times \exp \left( -\frac{A_{23}}{\left\vert \mathbf{a}\right\vert }%
\right) \exp \left( -\frac{A_{56}}{\left\vert \mathbf{b}\right\vert }\right)
d\mathbf{a}d\mathbf{b.}  \notag
\end{align}%
If one of the particles 5 and 6 is a neutron then $J=J_{n\beta }$, 
\begin{equation}
J_{n\beta }=A_{23}J_{3}  \label{JJ3}
\end{equation}%
with%
\begin{equation}
J_{3}=\int \int \frac{\delta \left[ \varepsilon \left( \mathbf{a},\mathbf{b}%
\right) \right] }{\left\vert \mathbf{a}\right\vert ^{5}\varepsilon
_{1f}^{2}\left( \mathbf{a}\right) }\exp \left( -\frac{A_{23}}{\left\vert 
\mathbf{a}\right\vert }\right) d\mathbf{a}d\mathbf{b}.  \label{J3}
\end{equation}%
Here and above%
\begin{equation}
\delta \left[ \varepsilon \left( \mathbf{a},\mathbf{b}\right) \right]
=\delta \left( \frac{\Delta }{8\mu _{56}c^{2}}b^{2}+\sqrt{a^{2}+\delta
_{e}^{2}}-1-\delta _{e}\right) .  \label{deltaab}
\end{equation}%
In obtaining $\left( \ref{deltaab}\right) $ the $\Delta a^{2}/\left( 8\mu
_{56}c^{2}\right) $ and $\Delta \left( \frac{1}{4m_{5}c^{2}}-\frac{1}{%
4m_{6}c^{2}}\right) \mathbf{a}\cdot \mathbf{b}$ terms in the argument can be
neglected since $\Delta /\left( \mu _{56}c^{2}\right) \ll 1$, $a\lesssim 1$,
and the dominant range in the integrals is where $b\gg a$. Using the $\delta %
\left[ g\left( x\right) \right] =\delta \left( x-x_{1}\right) /g^{\prime
}\left( x_{1}\right) $ identity, where $x_{1}$ is the root of the equation $%
g\left( x\right) =0$, the integrals $J_{2}$ and $J_{3}$ can be written as 
\begin{equation}
J_{2}=\beta _{2}J_{22}  \label{JJJ2}
\end{equation}%
with%
\begin{equation}
\beta _{2}=64\pi ^{2}\mu _{56}c^{2}/\Delta ,  \label{beta2}
\end{equation}%
\begin{equation}
J_{22}=\int_{0}^{a_{\max }}\frac{\exp \left( -\frac{A_{23}}{a}\right) \exp
\left( -\frac{A_{56}\sqrt{\Delta /\left( 8\mu _{56}c^{2}\right) }}{\sqrt{%
1+\delta _{e}-\sqrt{a^{2}+\delta _{e}^{2}}}}\right) da}{a^{3}\left( \sqrt{%
a^{2}+\delta _{e}^{2}}-\delta _{e}\right) ^{2}}  \label{J22}
\end{equation}%
and 
\begin{equation}
J_{3}=\beta _{3}J_{32}  \label{JJJ3}
\end{equation}%
with%
\begin{equation}
\beta _{3}=512\pi ^{2}\left( \mu _{56}c^{2}/\Delta \right) ^{2},
\label{beta3}
\end{equation}%
\begin{equation}
J_{32}=\int_{0}^{a_{\max }}\frac{\exp \left( -\frac{A_{23}}{a}\right) \left(
1+\delta _{e}-\sqrt{a^{2}+\delta _{e}^{2}}\right) da}{a^{3}\left( \sqrt{%
a^{2}+\delta _{e}^{2}}-\delta _{e}\right) ^{2}}  \label{J32}
\end{equation}%
where $a_{\max }=\sqrt{1+2\delta _{e}}$. Using all the above results, the
total rate $W_{tot}^{(2)}\left( +\beta \right) $ of the process having two
charged, heavy products reads 
\begin{equation}
W_{tot}^{(2)}\left( +\beta \right) =K_{tot}\left( \alpha \right)
K_{2}K_{+}\left\langle G_{S}\left( k_{1i},2k_{1f}\right) \right\rangle _{av}%
\frac{J_{22}}{\Delta ^{4}}h_{corr,3}^{2}uN_{2},  \label{Wtot+}
\end{equation}%
where%
\begin{equation}
K_{2}=z_{2}z_{3}\frac{512}{3\left( 1-\frac{2}{e}\right) ^{2}}\pi \left( 
\frac{R}{\hbar c}\right) ^{3}\alpha _{f}\mu _{23}c^{2}\left( \mu
_{56}c^{2}\right) ^{2}  \label{K2}
\end{equation}%
and%
\begin{equation}
K_{+}=\frac{\pi }{4}\alpha _{f}z_{5}z_{6}.  \label{K+}
\end{equation}%
The total rate $W_{tot}^{(2)}\left( n\beta \right) $ of the process, in
which one of the two heavy products is a neutron, reads%
\begin{equation}
W_{tot}^{(2)}\left( n\beta \right) =K_{tot}\left( \alpha \right)
K_{2}\left\langle G_{S}\left( k_{1i},2k_{1f}\right) \right\rangle _{av}\frac{%
J_{32}}{\Delta ^{4}}h_{corr,3}^{2}uN_{2}.  \label{Wtotn}
\end{equation}%
In order to compare the total rates $\left( \ref{Wtot0}\right) $, $\left( %
\ref{Wtot+}\right) $ and $\left( \ref{Wtotn}\right) $ of the processes of
different type, the quantities $F_{23}(\frac{\Delta }{\hbar c})$, $%
K_{2}K_{+}J_{22}$ and $K_{2}J_{32}$ have to be compared. The effect of
factor $\left\langle G_{S}\left( k_{1i},2k_{1f}\right) \right\rangle _{av}$
is not essential in the electron assisted processes.

\section{Appendix II. - Rate calculation of heavy particle assisted nuclear
processes}

In the case of heavy particle assisted nuclear processes the quantity $%
T_{if}^{(2)}$ modifies as 
\begin{align}
T_{if}^{(2)}(type)& =B(type)z_{1}z_{2}\delta \left( \mathbf{k}_{1f}+\mathbf{K%
}_{\left( type\right) }\right) \times  \label{Tfiheavybet} \\
& \int \frac{g_{S}f_{23}(k_{2\mu })}{\left( E_{\mu }-E_{i}\right) }\frac{%
\widetilde{\psi }_{1i}\left( \mathbf{k}_{2i}-\mathbf{k}_{1f}-\mathbf{k}%
_{2\mu }\right) }{\left\vert \mathbf{k}_{2i}-\mathbf{k}_{2\mu }\right\vert
^{2}+\lambda ^{2}}d\mathbf{k}_{2\mu }  \notag
\end{align}%
where%
\begin{equation}
E_{\mu }=\frac{\hbar ^{2}}{2m_{1}}k_{1\mu }^{2}+\frac{\hbar ^{2}}{2m_{2}}%
k_{2\mu }^{2},  \label{Emu}
\end{equation}

\begin{equation}
B(\alpha )=B_{0}\frac{\sqrt{3\pi R}}{V^{2}}
\end{equation}%
with 
\begin{equation}
B_{0}=-16\pi \alpha _{f}\left( 1-\frac{2}{e}\right) \left( \frac{f^{2}}{%
\hbar c}\right) \left( \hbar c\right) ^{2},
\end{equation}%
\begin{equation}
B(n\beta )=\frac{B(\alpha )}{\left( 1-\frac{2}{e}\right) }\sqrt{\frac{4\pi
R^{3}}{3V}}
\end{equation}%
and 
\begin{equation}
B(+\beta )=B(n\beta )f_{56}.
\end{equation}%
Since $\beta _{1}$ $\ll \left\vert \mathbf{K}\right\vert \approx \left\vert 
\mathbf{k}_{1f}\right\vert $, the Fourier transform of $\left( \ref{dfi}%
\right) $ 
\begin{equation}
\widetilde{\psi }_{1i}(\mathbf{K})=\frac{2^{3/2}\pi ^{3/4}}{\beta _{1}^{3/2}}%
e^{-\frac{\mathbf{K}^{2}}{2\beta _{1}^{2}}}  \label{a2k}
\end{equation}%
allows the approximation 
\begin{equation}
\widetilde{\psi }_{1i}(\mathbf{K})=8\pi ^{9/4}\beta _{1}^{3/2}\delta (%
\mathbf{K})  \label{Pszikal}
\end{equation}%
in $\left( \ref{Tfiheavybet}\right) $. As a result the integral over $%
\mathbf{k}_{2\mu }$ in $\left( \ref{Tfiheavybet}\right) $ can be carried
out, while $\left\vert \mathbf{k}_{2i}\right\vert $ and $\lambda $ can be
neglected since $\left\vert \mathbf{k}_{2i}\right\vert \ll \left\vert 
\mathbf{k}_{1f}\right\vert $ and $\lambda \ll \left\vert \mathbf{k}%
_{1f}\right\vert $. Neglecting $E_{i}$ in the denominator, $E_{\mu
}-E_{i}=\hbar ^{2}k_{1f}^{2}/\left( 2\mu _{12}\right) $. The result for $%
T_{if}^{(2)}(type)$ obtained in this way is 
\begin{align}
T_{if}^{(2)}(type)& =B(type)z_{1}z_{2}\delta \left( \mathbf{k}_{1f}+\mathbf{K%
}_{\left( type\right) }\right) \times   \label{TFi3} \\
& g_{S}\left( k_{2i},2k_{1f}\right) \mu _{12}\frac{f_{23}(k_{1f})}{\hbar
^{2}k_{1f}^{2}}\frac{16\pi ^{9/4}\beta _{1}^{3/2}}{k_{1f}^{2}}.  \notag
\end{align}

\subsection{Rates of heavy particle assisted nuclear capture processes}

First the process of $type=\alpha $ is dealt with. Substituting $\left( \ref%
{TFi3}\right) $ into $\left( \ref{Wfie}\right) $ the $\sum_{f}\rightarrow
\int \left[ V/\left( 2\pi \right) ^{3}\right] d\mathbf{k}_{4}\times \int %
\left[ V/\left( 2\pi \right) ^{3}\right] d\mathbf{k}_{1f}$ correspondence,
the identities $\left[ \delta \left( \mathbf{k}_{1f}+\mathbf{k}_{4}\right) %
\right] ^{2}=\delta \left( \mathbf{k}_{1f}+\mathbf{k}_{4}\right) \delta
\left( \mathbf{0}\right) $ and $\left( 2\pi \right) ^{3}\delta \left( 
\mathbf{0}\right) =V$ are used. When integrating over $\mathbf{k}_{4}$, the $%
\delta \left( \mathbf{k}_{1f}+\mathbf{k}_{4}\right) $ leads to the $\mathbf{k%
}_{4}=-\mathbf{k}_{1f}$ replacement. Therefore $E_{f}=E_{1f}\left(
k_{1f}\right) +E_{4}\left( k_{1f}\right) =\hbar ^{2}k_{1f}^{2}/\left( 2\mu
_{14}\right) $ in the energy Dirac delta. Applying the $k_{1f}dk_{1f}=\mu
_{14}dE_{f}/\hbar ^{2}$ and $k_{1f}=\sqrt{2\mu _{14}E_{f}}/\hbar $ relations
the integral over $k_{1f}$ is converted to an integral over $E_{f}$ and it
is carried out with the aid of the energy Dirac delta. Thus the transition
rate of a heavy particle assisted reaction of $type=\alpha $ 
\begin{align}
W_{fi}^{(2),h}\left( \alpha \right) & =\frac{K_{0}^{h}\left( \alpha \right)
\beta _{1}^{3}}{\sqrt{m_{0}c^{2}}}G_{S}\left( k_{2i},2k_{1f}\right)
h_{corr,3}^{2}\times  \label{Wfiha} \\
& \times \left( \frac{a_{12}}{a_{14}}\right) ^{2}\frac{\left(
z_{1}z_{2}\right) ^{2}}{\sqrt{a_{14}}}\frac{F_{23}^{h}(\Delta )}{\Delta
^{7/2}}\frac{1}{V}.  \notag
\end{align}%
For $K_{0}^{h}\left( \alpha \right) $ see $\left( \ref{K0halfa}\right) $,
the quantity $F_{23}^{h}(\Delta )$ is given by $\left( \ref{Fh230}\right) $,
the parameter $\eta _{23}^{h}\left( \Delta \right) $ is determined by $%
\left( \ref{eta23h0}\right) $, $m_{0}c^{2}=931.494$ $\left[ MeV\right] $ is
the atomic mass unit, and $a_{jk}$ is the reduced mass number of particles $%
j $ and $k$ of mass numbers $A_{j}$ and $A_{k}$ given by $\left( \ref{ajk}%
\right) $.

The total rate $W_{tot}^{(2),h}\left( \alpha \right) $ can be obtained
multiplying $W_{fi}^{(2),h}\left( \alpha \right) $ by $N_{2}$ and $N_{13}$. $%
N_{2}$ is the number of particles 2 and $N_{13}$ is the number of possible
pairs of particles 1 and 3. Introducing $n_{13}=N_{13}/V$, the number
density of particles 1 and 3, we take again $n_{13}=2u/v_{c}=8u/d^{3}$ valid
in the case of $fcc$ metals. 
\begin{align}
W_{tot}^{(2),h}\left( \alpha \right) & =\frac{8K_{0}^{h}\left( \alpha
\right) \beta _{1}^{3}}{d^{3}\sqrt{m_{0}c^{2}}}G_{S}\left(
k_{2i},2k_{1f}\right) h_{corr,3}^{2}\times  \label{Wfihatot} \\
& \times \left( \frac{a_{12}}{a_{14}}\right) ^{2}\frac{\left(
z_{1}z_{2}\right) ^{2}}{\sqrt{a_{14}}}\frac{F_{23}^{h}(\Delta )}{\Delta
^{7/2}}uN_{2}.  \notag
\end{align}

The parameter $\beta _{1}$ in the case of a localized deuteron is $\beta
_{1}^{d}=\sqrt{m_{d}\omega /\hbar }$, where $m_{d}$ is the deuteron rest
mass. In $Pd$, that is one of the possible materials which can absorb
hydrogen isotopes well, the energy of the ground state of the oscillator $%
E_{0}=\frac{3}{2}\hbar \omega $ $=72$ $\left[ meV\right] $\ \cite{Alefeld}
leading to $\hbar \omega =48$ $\left[ meV\right] $, that corresponds to the
energy of an optical phonon, and $\beta _{1}^{d}=4.81\times 10^{8}$ $\left[
cm^{-1}\right] $. This value alters as $\beta _{1}^{p}=4.04\times 10^{8}$ $%
\left[ cm^{-1}\right] $ for protons. With these numbers the characteristic
constants, which appear in $W_{tot}^{(2),h}\left( \alpha \right) $ $\left( %
\ref{Wfihatot}\right) $, are 
\begin{equation}
\frac{8K_{0}^{h}\left( \alpha \right) \left( \beta _{1}^{d}\right) ^{3}}{%
d^{3}\sqrt{m_{0}c^{2}}}=0.0031\left[ MeV^{7/2}s^{-1}\right]   \label{K0betdd}
\end{equation}%
and 
\begin{equation}
\frac{8K_{0}^{h}\left( \alpha \right) \left( \beta _{1}^{p}\right) ^{3}}{%
d^{3}\sqrt{m_{0}c^{2}}}=0.0018\text{ }\left[ MeV^{7/2}s^{-1}\right] 
\label{K0betdp}
\end{equation}%
in the case of deuterized and protonated $Pd$, and in the cases when
particle 1 is a deuteron or a proton, respectively.

\subsection{Rates of heavy particle assisted processes with two outgoing
heavy particles}

In the case of $type=+\beta $ $or$ $n\beta $ (for the process (b) of FIG. 2)
the calculation is modified as follows. Now the $\sum_{f}\rightarrow \int %
\left[ V/\left( 2\pi \right) ^{3}\right] d\mathbf{k}_{5}\times \int \left[
V/\left( 2\pi \right) ^{3}\right] d\mathbf{k}_{6}\times \int \left[ V/\left(
2\pi \right) ^{3}\right] d\mathbf{k}_{1f}$ correspondence and the identities 
$\left[ \delta \left( \mathbf{k}_{1f}+\mathbf{k}_{5}+\mathbf{k}_{6}\right) %
\right] ^{2}=\delta \left( \mathbf{k}_{1f}+\mathbf{k}_{5}+\mathbf{k}%
_{6}\right) \delta \left( \mathbf{0}\right) $ and $\left( 2\pi \right)
^{3}\delta \left( \mathbf{0}\right) =V$ are used and the integration over $%
\mathbf{k}_{1f}$ leads to the $\mathbf{k}_{1f}=-\mathbf{k}_{5}-\mathbf{k}_{6}
$ replacement. In the argument of the Dirac delta $E_{f}=E_{1f}\left(
\left\vert \mathbf{k}_{5}+\mathbf{k}_{6}\right\vert \right) +E_{5}\left(
k_{5}\right) +E_{6}\left( k_{6}\right) $, which can be rewritten as 
\begin{equation}
E_{f}=\frac{\hbar ^{2}k_{5}^{2}}{2\mu _{15}}+\frac{\hbar ^{2}k_{6}^{2}}{2\mu
_{16}}+\frac{\hbar ^{2}}{m_{1}}k_{5}k_{6}\zeta ,
\end{equation}%
where $\zeta =\cos \Theta _{56}$ with $\Theta _{56}$ the angle of vectors $%
\mathbf{k}_{5}$ and $\mathbf{k}_{6}$. Now new variables 
\begin{equation}
x_{j}=\sqrt{\frac{\hbar ^{2}}{2m_{1}\Delta }}k_{j},\text{ \ }j=5,6
\label{xj}
\end{equation}%
are introduced. With these variables 
\begin{equation}
E_{f}=\left( \frac{m_{1}}{\mu _{15}}x_{5}^{2}+\frac{m_{1}}{\mu _{16}}%
x_{6}^{2}+2x_{5}x_{6}\zeta \right) \Delta 
\end{equation}%
and 
\begin{equation}
k_{1f}^{2}=\frac{2m_{1}\Delta }{\hbar ^{2}}\left(
x_{5}^{2}+x_{6}^{2}+2x_{5}x_{6}\zeta \right) .
\end{equation}%
Applying the $\delta \left[ \chi \left( \zeta \right) \right] =\left\vert
\chi ^{\prime }\left( \zeta _{1}\right) \right\vert ^{-1}\delta \left( \zeta
-\zeta _{1}\right) $ identity, where $\zeta _{1}$ is the root of equation $%
\chi \left( \zeta \right) =0$, the integration over $\zeta $ can be carried
out. In our case 
\begin{equation}
\zeta _{1}=\frac{1-\left( \frac{m_{1}}{\mu _{15}}x_{5}^{2}+\frac{m_{1}}{\mu
_{16}}x_{6}^{2}\right) }{2x_{5}x_{6}}.
\end{equation}%
Introducing the notation 
\begin{equation}
g_{\left( \pm \right) }\left( x_{5},x_{6}\right) =x_{5}^{2}+x_{6}^{2}\pm
2x_{5}x_{6}\zeta _{1}
\end{equation}%
and after some algebra the transition rate 
\begin{align}
W_{fi}^{(2),h}\left( type\right) & =\frac{K_{0}\left( \beta \right) \beta
_{1}^{3}}{m_{1}c^{2}}G_{S}\left( k_{2i},2k_{1f}\right) h_{corr,3}^{2}\times 
\label{Wfitype} \\
& \times \left( z_{1}z_{2}\right) ^{2}\left( \frac{\mu _{12}c^{2}}{\Delta }%
\right) ^{2}J_{type}\frac{1}{V},  \notag
\end{align}%
where 
\begin{equation}
K_{0}\left( \beta \right) =2^{10}\pi ^{-1/2}\alpha _{f}^{2}\left( \frac{f^{2}%
}{\hbar c}\right) ^{2}R^{4}\hbar c^{2}.
\end{equation}%
For $type=+\beta $ the integral $J_{type}$ is 
\begin{equation}
J_{+\beta }=\int_{0}^{x_{6,up}}\int_{0}^{x_{5,up}}\frac{\Psi (\xi _{23})\Psi
(\xi _{56}^{-})}{g_{+}^{4}\left( x_{5},x_{6}\right) }x_{5}dx_{5}x_{6}dx_{6}
\end{equation}%
and for $type=n\beta $ the integral $J_{type}$ is%
\begin{equation}
J_{n\beta }=\int_{0}^{x_{6,up}}\int_{0}^{x_{5,up}}\frac{\Psi (\xi _{23})}{%
g_{+}^{4}\left( x_{5},x_{6}\right) }x_{5}dx_{5}x_{6}dx_{6}.
\end{equation}%
Here the following notation is used: 
\begin{equation}
\Psi (\xi )=\frac{\xi }{\exp \left( \xi \right) -1},
\end{equation}%
and the variables are 
\begin{equation}
\xi _{23}=2\pi z_{2}z_{3}\alpha _{f}\frac{\mu _{23}c}{\sqrt{2m_{1}\Delta }}%
g_{\left( +\right) }^{-1/2}\left( x_{5},x_{6}\right)   \label{kszi23}
\end{equation}%
and%
\begin{equation}
\xi _{56}^{-}=2\pi z_{5}z_{6}\alpha _{f}\frac{\mu _{56}c}{\sqrt{2m_{1}\Delta 
}}g_{\left( -\right) }^{-1/2}\left( x_{5},x_{6}\right)   \label{kszi56}
\end{equation}%
with 
\begin{equation}
g_{\left( \pm \right) }\left( x_{5},x_{6}\right) =x_{5}^{2}\left( 1\mp \frac{%
m_{1}}{\mu _{15}}\right) +x_{6}^{2}\left( 1\mp \frac{m_{1}}{\mu _{16}}%
\right) \pm 1.
\end{equation}%
The upper limits $x_{5,up}$ and $x_{6,up}$ are determined by the condition $%
\left\vert \zeta _{1}\right\vert \leq 1$.

Now the total rates $W_{tot}^{(2),h}\left( type\right) $ of reactions $%
type=+\beta $ $or$ $n\beta $ can also be obtained multiplying $%
W_{fi}^{(2),h}\left( type\right) $ by $N_{2}$ and $N_{13}$, and \ using $%
n_{13}=N_{13}/V=8u/d^{3}$ 
\begin{align}
W_{tot}^{(2),h}\left( type\right) & =\frac{8K_{0}\left( \beta \right) \beta
_{1}^{3}}{d^{3}m_{1}c^{2}}G_{S}\left( k_{2i},2k_{1f}\right)
h_{corr,3}^{2}\times  \label{Wfihtotb} \\
& \times \left( z_{1}z_{2}\right) ^{2}\left( \frac{\mu _{12}c^{2}}{\Delta }%
\right) ^{2}J_{type}uN_{2}.  \notag
\end{align}

In the rates $W_{tot}^{(2),h}\left( \alpha \right) $ $\left( \ref{Wfihatot}%
\right) $ and $W_{tot}^{(2),h}\left( type\right) $ $\left( \ref{Wfihtotb}%
\right) $ the function $G_{S}\left( k_{2i},2k_{1f}\right) =$ $%
F_{12}(k_{2i})/F_{12}(2k_{1f})$ plays important role since $%
F_{12}(k_{2i}(E_{i}))$ has strong energy dependence. If e.g. both particles
1 and 2 are protons then $2\pi \eta _{12}\left( E_{2i}\right) =0.700\times
E_{2i}^{-1/2}$ resulting $F_{12}(k_{2i}(E_{2i}))=6.4\times 10^{-3}$at $%
E_{2i}=0.01$ $\left[ MeV\right] $.

It is also an important aspect of the mechanism that if particle 1 is a
proton then in each proton assisted nuclear capture process an energetic
proton (of energy of about a few $MeV$) is created too. This proton in its
decelerating processes can create (secondary) free protons in the crystal,
if localized protons are also present. The secondary free protons may take
part in further proton assisted nuclear capture processes. Thus in this
process the secondary protons can play a role similar to that played by the
secondary neutrons in the case of nuclear fission.

\section{Appendix III. - Proton assisted proton capture processes}

Here some cases of the proton assisted proton captures 
\begin{equation}
\text{ }_{Z}^{A}X+p\rightarrow \text{ }_{Z+1}^{A+1}Y+\Delta
\end{equation}%
are investigated. Particles 1 and 2 are protons. Particles 3 and 4 have mass
numbers $A$ and $A+1$, respectively (see FIG. 2). As a first example we take
protonated $Ni$. In this case the possible processes are%
\begin{equation}
\text{ }_{28}^{A}Ni+p\rightarrow \text{ }_{29}^{A+1}Cu+\Delta .  \label{NiAp}
\end{equation}%
Most of the daughter nuclei $_{29}^{A+1}Cu$ decay by the 
\begin{equation}
\text{ }_{29}^{A+1}Cu+e\rightarrow \text{ }_{28}^{A+1}Ni+Q_{EC}
\label{CuA1e}
\end{equation}%
electron capture reaction. TABLE I. contains the relevant data for reactions 
$\left( \ref{NiAp}\right) $ and $\left( \ref{CuA1e}\right) $.

\begin{table}[ptb]
\tabskip=8pt 
\centerline {\vbox{\halign{\strut $#$\hfil&\hfil$#$\hfil&\hfil$#$
\hfil&\hfil$#$\hfil&\hfil$#$\hfil&\hfil$#$\hfil&\hfil$#$\cr
\noalign{\vskip2pt\hrule\vskip2pt}
\noalign{\hrule\vskip2pt\hrule\vskip2pt}
A &58 &59 &60 &61 &62 &64 \cr
\Delta ($MeV$) &3.417 &4.479 &4.801 &5.867 &6.122 &7.452 \cr
r_{A} &0.68077 &(2.4\times10^{12}$s$) &0.26223 &0.0114 &0.03634 &0.00926 \cr
2\pi \eta _{23}^{h} &15.04 &13.14 &12.69 &11.48 &11.24 &10.18 \cr
s_{A}\left( \Delta \right)\times10^{8} ($MeV$^{-7/2}) &4.08 &- &4.23 &0.276 &0.945 &0.316 \cr
\tau (s) &81.5 &1422 &1.20\times10^{4} &584.4 &$stable$ &$stable$ \cr
Q_{EC} ($MeV$) &4.789 &6.127 &2.237 &3.948 &$-$ &$-$ \cr
\noalign{\vskip2pt\hrule\vskip2pt\hrule}}}}
\par
\vskip20pt
\par
\vskip50pt
\caption{Numerical data of the $_{28}^{A}Ni+p\rightarrow\text{ }%
_{29}^{A+1}Cu+\Delta$ and $_{29}^{A+1}Cu+e\rightarrow\text{ }%
_{28}^{A+1}Ni+Q_{EC}$ reactions. Computed values of $2\protect\pi\protect\eta%
_{23}^{h}$ and $s_{A}\left( \Delta\right) $ are given (for their formula see
the text). $A$ is the mass number, $r_{A}$ is the relative natural
abundance, $\protect\tau$ is the half life of the $_{29}^{A+1}Cu$ isotope
produced. It decays by electron capture of reaction energy $Q_{EC}$. In the
row $r_{A}$ at $A=59$ the $2.4\times10^{12}$ $s$ stands for the life time of
the unstable $_{28}^{59}Ni$. }
\label{Table1}
\end{table}

\begin{table}[tbp]
\tabskip=8pt 
\centerline {\vbox{\halign{\strut $#$\hfil&\hfil$#$\hfil&\hfil$#$
\hfil&\hfil$#$\hfil&\hfil$#$\hfil&\hfil$#$\hfil&\hfil$#$\cr
\noalign{\vskip2pt\hrule\vskip2pt}
\noalign{\hrule\vskip2pt\hrule\vskip2pt}
A &102 &104 &105 &106 &108 &110 \cr
\Delta ($MeV$) &4.155 &4.966 &5.814 &5.789 &6.487 &7.156 \cr
r_{A} &0.0102 &0.1114 &0.2233 &0.2733 &0.2646 &0.1172 \cr
2\pi \eta _{23}^{h} &22.41 &20.49 &18,94 &18.98 &17.93 &17.07 \cr
s_{A}\left( \Delta \right)\times10^{11} ($MeV$^{-7/2}) &0.0285 &1.06 &5.31 &6.35 &11.1 &7.80 \cr
\tau (s) &4002 &3.56\times10^{6} &1438 &$stable$ &$stable$ &6.44\times10^{5} \cr
Q_{EC} ($MeV$) &2.688 &1.346 &2.965 &$-$ &$-$ &$no EC$ \cr
\noalign{\vskip2pt\hrule\vskip2pt\hrule}}}}
\par
\vskip20pt
\par
\vskip50pt
\caption{Numerical data of the $_{46}^{A}Pd+p\rightarrow \text{ }%
_{47}^{A+1}Ag+\Delta $ and $_{47}^{A+1}Ag+e\rightarrow \text{ }%
_{46}^{A+1}Pd+Q_{EC}$ reactions. Computed values of $2\protect\pi \protect%
\eta _{23}^{h}$ and $s_{A}\left( \Delta \right) $ are given (for their
formula see the text). $A$ is the mass number, $r_{A}$ is the relative
natural abundance, $\protect\tau $ is the half life of the $_{47}^{A+1}Ag$
isotope produced. It decays by electron capture of reaction energy $Q_{EC}$. 
}
\label{Table2}
\end{table}

An other interesting family of the heavy particle (proton) assisted proton
capture is 
\begin{equation}
\text{ }_{46}^{A}Pd+p\rightarrow \text{ }_{47}^{A+1}Ag+\Delta ,  \label{PdAp}
\end{equation}%
that is mainly followed by the 
\begin{equation}
\text{ }_{47}^{A+1}Ag+e\rightarrow \text{ }_{46}^{A+1}Pd+Q_{EC}
\label{AgA1e}
\end{equation}%
reaction. (The relevant data can be found in TABLE II. The nuclear data of
TABLES I. and II. are taken from \cite{Shir}.)

For the processes $\left( \ref{NiAp}\right) $ and $\left( \ref{PdAp}\right) $
$a_{12}=1/2$ and $a_{14}=a_{23}=1$ is a good approximation in the case of
protonated $Ni$ and $Pd$. First we take a protonated $Ni$ $\left(
z_{3}=28\right) $. The rate $W_{tot}^{(2),h}\left( \alpha ,A\right) $
belonging to process $\left( \ref{NiAp}\right) $ is 
\begin{equation}
W_{tot}^{(2),h}\left( \alpha ,A\right) =G_{S}\frac{2K_{0}^{h}\left( \alpha
\right) \left( \beta _{1}^{p}\right) ^{3}}{d^{3}\sqrt{m_{0}c^{2}}}\left(
A-z_{3}\right) ^{2}s_{A}\left( \Delta \right) uN_{2}.  \label{WhtotA}
\end{equation}%
Here $K_{0}^{h}\left( \alpha \right) $ is given by $\left( \ref{K0halfa}%
\right) $, $d$ is the lattice parameter, $m_{0}c^{2}=931.494$ $\left[ MeV%
\right] $ is the atomic mass unit, and for $\beta _{1}^{p}$ see Appendix II.
A under $\left( \ref{Wfihatot}\right) $. These all result $2K_{0}^{h}\left(
\alpha \right) \left( \beta _{1}^{p}\right) ^{3}d^{-3}\left(
m_{0}c^{2}\right) ^{-1/2}=4.6\times 10^{-4}\left[ MeV^{7/2}s^{-1}\right] $. $%
N_{2}$ is the actual number of free protons in the sample, 
\begin{equation}
s_{A}\left( \Delta \right) =r_{A}\frac{F_{23}^{h}(\Delta )}{\Delta ^{7/2}},
\label{sAdelta}
\end{equation}%
where $r_{A}$ is the relative natural abundance and%
\begin{equation}
2\pi \eta _{23}^{h}=\frac{27.8\text{ }\left[ MeV^{1/2}\right] }{\sqrt{\Delta 
}}  \label{eta23hA}
\end{equation}%
with $z_{3}=28$ in $\left( \ref{eta23h0}\right) $. The formula $\left( \ref%
{WhtotA}\right) $ can be also used in the case of processes $\left( \ref%
{PdAp}\right) $ with $s_{A}$ and $\Delta $ data of TABLE II. and using 
\begin{equation}
2\pi \eta _{23}^{h}=\frac{45.7\text{ }\left[ MeV^{1/2}\right] }{\sqrt{\Delta 
}},  \label{etha23hPd}
\end{equation}%
that comes from $\left( \ref{eta23h0}\right) $ with $z_{3}=46$
(corresponding to $Pd$).

The processes $\left( \ref{NiAp}\right) \ $and $\left( \ref{PdAp}\right) $
may produce stable $_{29}^{62}Cu$, $_{29}^{64}Cu$ and $_{47}^{106}Ag$, $%
_{47}^{108}Ag$ isotopes, respectively, whose heavy particle assisted proton
capture reaction may give rise to a chain of nuclear transmutations.

\end{document}